\documentclass{jetpl}
\twocolumn
\title{Topology of chiral superfluid: skyrmions, Weyl fermions and chiral anomaly
\\
{\it devoted to Vladimir Mineev's 70'th jubilee}}
\rtitle{Topology of chiral superfluid: skyrmions, Weyl fermions and chiral anomaly}

\sodtitle{Topology of chiral superfluid: skyrmions, Weyl fermions and chiral anomaly}

\author
{G.E. Volovik $^{+\#}$ \/\thanks{e-mail: volovik@boojum.hut.fi}
}

\rauthor{
G.E.Volovik
}

\sodauthor{
Volovik
}

\address{
$^{+}$ Low Temperature Laboratory, Department of Applied Physics, Aalto University, PO Box 15100, FI-00076 AALTO, Finland
\\
$^{\#}$ Landau Institute for Theoretical Physics RAS, Kosygina 2, 119334 Moscow, Russia
}

\abstract{ 
Chiral anomaly observed in the chiral superfluid $^3$He-A is the result of the combined effect of the real space and momentum space topologies. This effect incorporates several topological charges in the extended $({\bf k},{\bf r})$-space, which is beyond the conventional chiral anomaly in the relativistic systems. 
}

\begin{document}

\maketitle

Discovery of superfluid phases of $^3$He in 1972\cite{OsheroffRichardsonLee1972} with
their multi-component order parameter opened the new area of the application of topology to condensed matter systems. In addition to many exotic topological defects (some of them have been observed in these phases, the others are still waiting for their creation and detection), the momentum-space topology plays an important role in the properties of these phases.
Probably the most interesting topological object in momentum space is the Weyl node in the quasiparticle spectrum. Close to the Weyl points, quasiparticles obey the Weyl equation and behave as Weyl fermions, with all the accompanying effects such as chiral anomaly described by the Adler-Bell-Jackiw equation.\cite{Adler1969,Adler2005,BellJackiw1969}

The Weyl points exist in the fermionic spectrum of superfluid $^3$He-A and also
in the core of quantized vortices in $^3$He-B.\cite{1982}
 Discovery of Weyl fermions has been also reported  in the topological semiconductors.\cite{Weng2015,Huang2015,Lv2015,Lu2015} 
Within the topological defects (vortices, skyrmions, solitons, etc.)  the position ${\bf K}_a$ of $a$-th  Weyl points in momentum ${\bf k}$-space depends on the coordinates ${\bf r}$.
Such dependence leads to the effective (synthetic) electric and magnetic fields acting on the Weyl quasiparticles. The topological protection of the Weyl points together with topology of the spatial distribution of the Weyl points in the coordinate space gives rise to the more complicated combined topology in the extended phase space $({\bf k},{\bf r})$.\cite{1982} This combined topology connects the effect of chiral anomaly and the dynamics of skyrmions, which allowed us to observe experimentally the consequence of the chiral anomaly and to verify the Adler-Bell-Jackiw equation.\cite{BevanNature1997} 

Let us start with the gap function for the spin-triplet $p$-wave pairing in superfluid phases 
of $^3$He. In the
representation $S=1$ ($S$ is the spin momentum of Cooper pairs) and $L=1$
($L$ is the orbital angular momentum of Cooper pairs) the gap function depends linearly on spin
${\mbox{\boldmath$\sigma$}}$ and momentum ${\bf k}$. Thus it can be represented in terms of the $3\times 3$ matrix  order parameter $A_{\alpha i}$, which in inhomogeneous case depends on the coordinate ${\bf r}$
\begin{equation}
\hat\Delta({\bf k})=A_{\alpha i}  \sigma_\alpha i\sigma_y \frac{k_i}{k_F} ~.
\label{Triplet}
\end{equation}
The corresponding Bogoliubov-de Gennes Hamiltonian density is
\begin{equation}
H({\bf k})=  
\begin{pmatrix} 
\epsilon({\bf k}) & \hat\Delta({\bf k})
\\ 
\hat\Delta^*({\bf k}) & -\epsilon({\bf k})
\end{pmatrix}  \,, \,\epsilon({\bf k})=\frac{k^2 -k_F^2}{2m} \,.
\label{H}
\end{equation}

In bulk liquid $^3$He there are two topologically different phases:  the chiral superfluid $^3$He-A with topologically protected Weyl points in the quasiparticle spectrum 
and the fully gapped  time reversal invariant superfluid $^3$He-B (on the momentum space 
topology in superfluid $^3$He-B, which leads to Majorana fermions living on the surface, see recent review \cite{Mizushima2015}.   It is instructive to consider the path between the two phases to see how the topology changes. In Ref. \cite{1982} the following Ansatz has been discussed with the  parameter $a$ changing from $a=0$ in $^3$He-A to $a=1$ in  $^3$He-B: 
\begin{equation}
\hat\Delta({\bf k})= \frac{\Delta}{k_F}
\begin{pmatrix} 
 -k_x +(2a-1)ik_y & ak_z
\\ 
ak_z & k_x+ik_y 
\end{pmatrix} \,.
\label{AtoB}
\end{equation}
For $a=1$ one has a nodeless isotropic state:
\begin{equation}
\hat\Delta= \frac{\Delta}{k_F}
\begin{pmatrix} 
 -k_x +ik_y &k_z
\\ 
k_z & k_x+ik_y 
\end{pmatrix}
=   \frac{\Delta}{k_F} ( {\mbox{\boldmath$\sigma$}}  \cdot \bf k) i\sigma_y \,,
\label{B}
\end{equation}
while for $a=0$ the gap function describes the chiral superfluid with two Weyl points
at ${\bf K}_\pm=\pm k_F\hat{\bf z}$:
\begin{equation}
\hat\Delta= \frac{\Delta}{k_F}
\begin{pmatrix} 
 -k_x -ik_y &0
\\ 
0 & k_x+ik_y 
\end{pmatrix}
=   \frac{\Delta}{k_F} \sigma_x i\sigma_y (k_x + ik_y)\,.
\label{A}
\end{equation}

In general, the vacuum manifold in a chiral $p$-wave
superfluid ($^3$He-A) is represented in terms of the vector in spin space and the triad in the orbital space:
\begin{equation}
A_{\alpha i}= \Delta \hat d_\alpha( \hat m_i + i\hat n_i) ~.
\label{OrderParameterA}
\end{equation}
Here $\hat{\bf d}$ is the unit vector of the spin-space anisotropy; in Eq.(\ref{A}) $\hat{\bf d}$ is oriented along $\hat{\bf x}$.
Vectors $\hat{\bf m}$ and  
$\hat{\bf n}$ are mutually orthogonal  unit vectors in the orbital space, which together with the orbital momentum vector  
$\hat{\bf l}=\hat{\bf m}\times\hat{\bf n}$ form the orbital triad. 
 In Eq.(\ref{A}) $\hat{\bf l}$ is oriented along $\hat{\bf z}$.
The vectors  $\hat{\bf m}$ and  $\hat{\bf n}$  
determine the superfluid velocity of the chiral condensate,
${\bf v}_{\rm s}=\frac{\hbar}{2m}\hat m_i\nabla \hat n_i$, where $m$ is the mass of the $^3$He atom. That is why the texture of the unit vector $\hat{\bf l}$ carries vorticity according to the Mermi-Ho relation\cite{MerminHo1976}:
\begin{equation}
  \nabla\times{\bf v}_{\rm s} =
      {\hbar\over 4m~}e_{ijk} \hat l_i{\bf \nabla} \hat l_j\times{\bf
\nabla}
\hat l_k \,.
\label{Mermin-HoEq}
\end{equation}

The classification of the topological objects in the fields $\hat{\bf d}$ and 
$\hat{\bf m}+i\hat{\bf n}$ made in Refs.\cite{1976b,1977a,1977b,1978a} revealed the possibility of many configurations with nontrivial topology. Among them there are: vortex-skyrmion -- the continuous  $\hat{\bf l}$-texture which according to Eq.(\ref{Mermin-HoEq}) represents the doubly quantized vortex\cite{Chechetkin1976,AndersonToulouse1977} (observed in rotating cryostat\cite{Seppala1984}); topological solitons described by relative 
homotopy groups\cite{1977b,1978a}  (vorticity concentrated in the core of the topological soliton forms the vortex sheet  observed under rotation in Refs.\cite{Parts1994a,Parts1994b}); the half-quantum vortex\cite{1976b} -- the condensed matter analog of the Alice string in particle physics\cite{Schwarz1982} (experimentally it has been stabilized and detected only recently in the polar phase of $^3$He\cite{Autti2015}); vortex terminated by hedgehog\cite{Blaha1976,1976a}  -- the condensed matter analog of electroweak magnetic monopole and the other monopoles connected by strings\cite{Kibble2008} (experimentally it was found in cold gases\cite{Mottonen2014}); etc.
Among different topological objects observed in superfluid $^3$He-B we discuss here the 
axisymmetric vortex with the spontaneously broken parity, which exists  in $^3$He-B at high pressure.\cite{1983}
At lower pressure the first order phase transition with the spontaneous symmetry breaking in the vortex core takes place\cite{Kondo1991}, which corresponds to the splitting of the vortex core into two half-quantum vortices (see the latest reference \cite{Silaev2015}).

In $^3$He-A, the orbital vector $\hat{\bf l}$ also determines the position of the Weyl points in the fermionic spectrum of the Hamiltonian (\ref{H}). The Hamiltonian is nullified when $k=k_F$, i.e. on the former Fermi surface, and when the determinant of the gap function is nullified, i.e. when ${\bf k}\cdot \hat{\bf m}={\bf k}\cdot \hat{\bf n}=0$. Altogether this gives two Weyl points at ${\bf K}_\pm=\pm k_F\hat{\bf l}$.

The isolated Weyl point  is protected by topological invariant and survives when the interaction between quasiparticles is taken into account.  In the modern language the integer valued topological invariant $N$ for the Weyl point is determined as the magnetic charge 
of the Berry phase monopole.\cite{Volovik2003} Originally this invariant has been calculated in a different way, see Ref.\cite{1982}. 
As follows from Eq.(\ref{H}) the node in the spectrum occurs when $\epsilon({\bf  k})=0$ and  the determinant of the gap function is zero, ${\rm det}\,\hat\Delta({\bf k})=0$. That is why the node represents the crossing point of the nodal surface $\epsilon({\bf  k})=0$ (the Fermi surface of the original normal state of liquid $^3$He) and the nodal line  where ${\rm det}\,\hat\Delta({\bf k})=0$. The latter is described by the integer winding number of the  phase $\Phi({\bf k})$ of the determinant:
\begin{equation}
{\rm det}\,\hat\Delta({\bf k})=|{\rm det}\,\hat\Delta({\bf k})| e^{i\Phi({\bf k})} \,,
\label{Determinant}
\end{equation}
 around the nodal line.
The topological charge $N$ is the winding number of the phase $\Phi({\bf k})$  when viewed from the region outside the Fermi surface.\cite{1982}
Such definition of the topological charge $N$ coincides with the Berry monopole charge, giving $N= \pm 2$ for the Weyl points at ${\bf K}_\pm=\pm k_F\hat{\bf l}$.

In general, the state with the topological charge larger than unity, $|N|>1$, is unstable towards splitting to the states with nodes described elementary charges $N=\pm 1$. In $^3$He-A the nodes with topological charges $N= \pm 2$ are protected by the discrete $Z_2$ symmetry related to spins. On the way from $^3$He-A to the $^3$He-B in Eq.(\ref{AtoB}),  the spin symmetry   is violated and  at $a>0$   the nodes with $|N|=2$ split into the nodes with $|N|=1$.

\begin{figure}
 \includegraphics[width=0.5\textwidth]{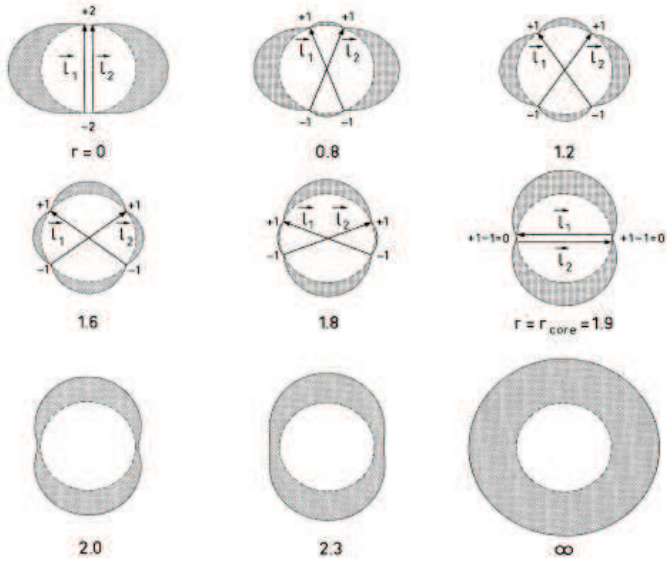}
 \caption{Fig. 1. Combined topology of the Weyl points in the  core of the axisymmetric $^3$He-B vortex
according to Ref. \cite{1982} (from review paper \cite{SV1987}). Here $r$ is the distance from the vortex axis. Weyl points are topologically stable nodes in the quasiparticle particle spectrum, 
which have integer topological charge  $N$ in momentum space. The  Weyl points are situated at momenta 
${\bf K}_a$  in momentum space, which depend on the position ${\bf r}$ in the real space.
In superfluid $^3$He, the Weyl points live at $k=k_F$, i.e. on the former Fermi surafce, ${\bf K}_a=\pm k_F\hat{\bf l}_a$, where $\hat{\bf l}_a$ are unit vectors. That is why originally the Weyl point was called "boojum on Fermi surface".   On the vortex axis, at $r=0$, one has two pairs of Weyl points with $\hat{\bf l}_1=\hat{\bf l}_2=\hat{\bf z}$. Each pair forms the Weyl point with double topological charges, $N=+2$ on the north pole and  $N=-2$ on the south pole.
This corresponds to the chiral $^3$He-A on the axis without any vorticity.
For $r>0$, the multiple nodes split into pairs of Weyl points, each carrying unit topological charges $N=+1$ or $N=-1$. For increasing $r$, the Weyl points move continuously towards the
equatorial plane, where they annihilate each other ($+1-1=0$). For larger $r$ the fully gapped state is formed, which becomes the isotropic $^3$He-B  far from the vortex.
The coordinate dependence of the Weyl point gives rise to vorticity concentrated in the vortex core, as a result the vortex in the B-phase acquires the winding number. In other words, according to Ref. \cite{1982} the vortex -- the topological defects  in ${\bf r}$-space --  flows out into ${\bf k}$-space due to evolution of the Weyl points. The topology of the evolution is governed by Eq.(\ref{Combined}), which connects three topological invariants: real-space winding number of the vortex ${\cal N}$, momentum-space invariant of the Weyl point $N$ and the invariant $\nu$, which describes the evolution of the Weyl point in real space.
 }
 \label{Evolution}
\end{figure}

The evolution of the  Weyl points takes place in the  core of vortices. In the core of the axisymmetric vortex with the spontaneously broken parity in $^3$He-B\cite{1983} one has the  $^3$He-A order parameter on the vortex axis, which continuously transforms to the  $^3$He-B order parameter far from the core. This means that corresponding parameter $a$ depends on the distance $r$ from the vortex axis, see Ref. \cite{1982} and numerical simulations in Fig. \ref{Evolution} from Ref. \cite{SV1987}. On the vortex axis one has $a(r=0)=0$, which corresponds to the chiral  $^3$He-A without vorticity, but with the doubly degenerate Weyl points in momentum space with topological charges $N=\pm 2$.  In the interval $0<r<r_{\rm core}$,   the Weyl points split into the elementary Weyl points with $N=\pm 1$.
In the Ansatz Eq.(\ref{AtoB}) this corresponds to the interval  $0<a<1/2$.  At $r=r_{\rm core}$ the Weyl points with opposite $N$  merge to form the Dirac points with trivial topological charge, $N=0$.  At $r>r_{\rm core}$, the Dirac points disappear, because they are not protected by topology,  and the fully gapped state emerges.  Far from the vortex core one obtains $a(r=\infty)=1$, which corresponds to the  $^3$He-B with the $2\pi$ phase winding around the vortex line.

In this evolution of Weyl points the  chirality of $^3$He-A, which is the property of topology in momentum space,  continuously transforms to the integer valued circulation of superfluid velocity around the vortex, which is described by the real space topology.
The topological connection of the real-space and momentum-space properties is encoded in equation (4.7) of Ref.\cite{1982}:
\begin{equation}
{\cal N} =\frac{1}{2}\sum_a N_a\nu_a \,.
\label{Combined}
\end{equation}
Here ${\cal N}$ is the real-space topological invariant -- the winding number of the vortex;
$N_a$ is the momentum-space topological invariant describing the $a$-th Weyl point.
Finally the index $\nu_a$ connects the two spaces: it shows how many times the Weyl point ${\bf K}_a$  covers sphere, when the coordinates ${\bf r}=(x,y)$ run over the cross-section of the vortex core:
\begin{equation}
\nu_a=\frac{1}{4\pi} \int dxdy \frac{1}{|{\bf K}^a|^3}{\bf K}^a\cdot(\partial_x {\bf K}^a \times \partial_y {\bf K}^a)\,.
\label{nu}
\end{equation}

For the discussed $^3$He-B vortex the Weyl nodes with $N_a=\pm 1$ cover the half a sphere,  $\nu_a=\pm 1/2$, which gives ${\cal N} =\frac{1}{2}(1/2 + 1/2+ 1/2+1/2)=1$.
 
\begin{figure}
 \includegraphics[width=0.5\textwidth]{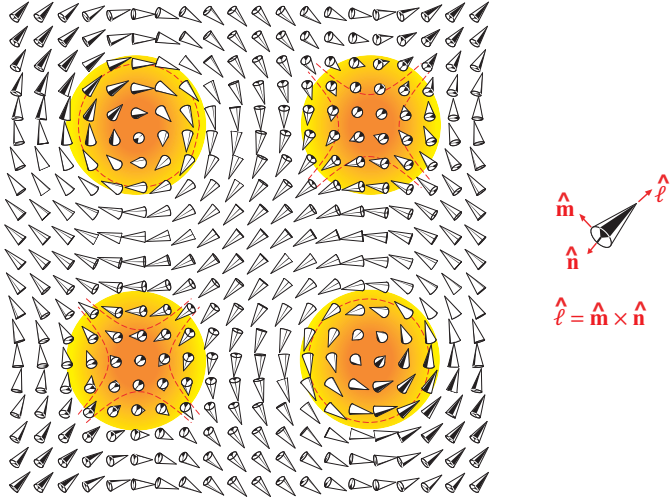}
 \caption{Fig.2. Elementary cell of meron lattice in rotating superfluid $^3$He-A at low magnetic field (adapted from \cite{Eltsov2000}). The unit cell  consists of four merons,  two of which are circular and two are hyperbolic. In each meron the vector $\hat{\bf l}$ and thus the Weyl points with $N_a=\pm 2$ cover half a sphere. According to
 Eq.(\ref{Combined}) each meron represents a vortex with a single quantum of circulation, 
${\cal N}=1$. Thus the unit cell carries topological charge ${\cal N}=4$.   In the circular vortex-meron the orbital vector $\hat{\bf l} \parallel \mathbf{\Omega}$ in the center, where $\mathbf{\Omega}$
is the angular velocity of the rotating cryostat.  In the hyperbolic vortex-meron $\hat{\bf l} \parallel - \mathbf{\Omega}$ in the center. 
At high magnetic field the rotating state is an array of isolated skyrmions with ${\cal N}=2$ each. The skyrmion contains two merons.
 }
 \label{Skyrmion}
\end{figure}

Eq.(\ref{Combined}) is also applicable to the continuous textures  in $^3$He-A.  For example, in the skyrmion\cite{Chechetkin1976,AndersonToulouse1977} the Weyl nodes with $N_a=\pm 2$ cover the whole sphere once,  $\nu_a=\pm 1$. This gives ${\cal N} =\frac{1}{2}(2\times 1 + (-2)\times(-1))=2$, which means that  the skyrmion represents the continuous doubly quantized vortex. The steps in the NMR spectrum corresponding to the ${\cal N}=2$ vortices were observed  in the NMR experiments on rotating $^3$He-A.\cite{Blaauwgeers2000}   The meron -- the Mermin-Ho vortex with $N_a=\pm 2$ and 
 $\nu_a=\pm 1/2$ -- represents the singly quantized vortex, ${\cal N} =\frac{1}{2}(2\times \frac{1}{2} + (-2)\times (-\frac{1}{2}))=1$. The elementary cell of the skyrmion lattice in the low magnetic field contains 4 merons, see Fig. \ref{Skyrmion}, and thus has ${\cal N}=4$. Merons are also the building blocks of the vortex sheet observed in $^3$He-A.\cite{Parts1994a,Parts1994b} 
 
 One can try to apply Eq.(\ref{Combined}) to the Alice string -- the half quantum vortex in $^3$He-A,\cite{1976b}
 assuming that as in the $^3$He-B the parity is spontaneously broken in the vortex core.
 Since the  half quantum vortex can be represented as the singly quantized vortex in one spin component, one can consider the evolution of the Weyl nodes with $N_a=\pm 1$ only in that component.  To get ${\cal N}=1/2$ these nodes should cover the half a sphere,  $\nu_a=\pm 1/2$, which is impossible in continuous way. This suggests that the core of the half quantum vortex should contain nodes of higher dimension -- such as the nodal line on the axis of the $^3$He-B 
 disclination.\cite{Grinevich1988} 
 
The close connection between topologies in real and momentum space can be also seen when equation (\ref{Determinant}) is extended to the inhomogeneous case
\begin{equation}
{\rm det}\,\hat\Delta({\bf k},{\bf r})=|{\rm det}\,\hat\Delta({\bf k},{\bf r})| e^{i\Phi({\bf k},{\bf r})} \,.
\label{DeterminantExtended}
\end{equation}
Then the phase $\Phi({\bf k},{\bf r})$ may include the winding in real space (a vortex) and the winding in momentum space, which gives rise to Weyl point.  In general the winding  number of the phase protects the 4-dimensional vortex in the 6-dimensional $({\bf k},{\bf r})$-space. By changing the orientation of this 4-D manifold of zeroes in the 6-D space, one can transform the $^3$He-B state with a vortex, where
$\Phi({\bf k},{\bf r})=\Phi({\bf r})=2\phi$, to the vortex-free homogeneous $^3$He-A state, where  $\Phi({\bf k},{\bf r})=\Phi({\bf k})$ has the winding number in momentum space.\cite{1982}

For the inhomogeneous $^3$He-A, the nonzero winding of the phase $\Phi({\bf k},{\bf r})$ gives rise to the following  4-D singularity  in the 6-D $({\bf k},{\bf r})$-space:
\begin{eqnarray}
\left(
 \frac{\partial}{\partial {\bf k}} \cdot  \frac{\partial}{\partial {\bf r}} -  \frac{\partial}{\partial {\bf r}} \cdot  \frac{\partial}{\partial {\bf k}}
 \right)  \Phi({\bf k},{\bf r})=
 \nonumber
 \\
 = -2\pi\, (\hat{\bf l}({\bf r}) \cdot {\bf k})\, (\hat{\bf l}({\bf r}) \cdot \nabla \times \hat{\bf l}({\bf r})) 
 \, \delta\left({\bf k}-\hat{\bf l}({\bf r}) (\hat{\bf l}({\bf r}) \cdot {\bf k})\right)
 \,.
\label{Singularity}
\end{eqnarray}
Such singularity leads to the anomalies in the equations for the mass current (linear momentum density) and angular momentum density of the chiral liquid, since these quantities can be expressed via the gradients of the generalized phase $\Phi({\bf k},{\bf r})$.\cite{1982} Later it became clear, that  these anomalies are the manifestation of the chiral anomaly in $^3$He-A related to the Weyl points. The effect of the chiral anomaly has been observed in experiments with dynamics of the 
vortex-skyrmions,\cite{BevanNature1997} which revealed the existence of the anomalous spectral-flow force acting on skyrmions. 

Originally the anomalies in the dynamics of $^3$He-A have been obtained from the calculations of the response functions and from the hydrodynamic equations, which take into account the singularity of the phase $\Phi$ 
in Eq.(\ref{Singularity}).\cite{1981,1982}  In these calculations the main contribution to the anomalous behavior comes from the momenta ${\bf k}$ far from the Weyl points, where the spectrum is highly nonrelativistic.
The results of calculations coincide with the later results obtained using the relativistic spectrum 
of chiral fermions emerging in the vicinity of  the Weyl points.  This is because the spectral flow through the Weyl nodes, which is in the origin of anomalies, does not depend on energy and is the same far from and close to the nodes.
The original approach gives the following anomalous contributions to the mass current density,
the momentum conservation law, and the angular momentum  conservation law at $T=0$:\cite{1981} 
\begin{eqnarray}
 {\bf j} = mn {\bf v}_{\rm s} + \frac{1}{2}\nabla \times \left( \frac{1}{2}  n \hat{\bf l} \right)
-\frac{1}{2}C_0 \hat{\bf l} (\hat{\bf l}\cdot \nabla \times \hat{\bf l}) 
\,,
   \label{Current}
   \\
\partial_t j_i -\nabla_k\pi_{ik} = -\frac{3}{2}C_0 \hat l_i (\partial_t\hat{\bf l}\cdot \nabla \times \hat{\bf l}) \,,
\label{nonconservation}
\\
\delta {\bf L} = \delta\left( \frac{1}{2}  n \hat{\bf l} \right) -\frac{1}{2}C_0 \delta \hat{\bf l} 
=\frac{1}{2}  (n -C_0) \delta\hat{\bf l}  -\frac{1}{2}\hat{\bf l}  \delta n \,.
\label{L}
\end{eqnarray}
Here $\hbar=1$; $n$ is the particle density; $\frac{\hbar}{2}  n \hat{\bf l}$ is what one expects for the density of angular momentum of the liquid with $n/2$ Cooper pairs, each with angular momentum $L=1$ along $\hat{\bf l}$. In Eq.(\ref{Current}),  the first two terms are correspondingly the superfluid current with velocity $ {\bf v}_{\rm s}$ and the current induced by the inhomogeneity of the angular momentum density. The extra contribution to the current contains the parameter $C_0$:
\begin{equation}
 C_0=\frac{k_F^3}{3\pi^2\hbar^2} \,.
\label{C0}
\end{equation}
The same parameter $C_0$ enters the rhs of Eq.(\ref{nonconservation}). The nonzero value of the rhs  of Eq.(\ref{nonconservation}) manifests the non-conservation of the vacuum current, which means that the linear momentum is carried away by the fermionic quasiparticles created from the superfluid vacuum. The Eq.(\ref{L}) shows the variation of the angular momentum $\delta {\bf L}$, which enters the dynamic equation $\partial_t \delta {\bf L}= - \delta E / \delta  {\mbox{\boldmath$\theta$}}$,
where $\delta  {\mbox{\boldmath$\theta$}}$ is the angle of the infinitesimal rotation of the orbital triad. The same parameter $C_0$ in this equation demonstates that 
the canonical equation for the density of the internal angular momentum of the liquid is non-local, and there is the dynamical reduction of the angular momentum from the static value $\frac{\hbar}{2}  n \hat{\bf l}$  to the dynamic value $\frac{\hbar}{2}  (n - C_0)\hat{\bf l}$. 
Note that in $^3$He-A the Cooper pairing is in the weak coupling regime, which means that the gap amplitude $\Delta$ is much smaller that the Fermi energy. As a result the particle density $n$ in the superfluid state is very close to the parameter $C_0$, which is equal to the particle density in the normal state, $C_0=n(\Delta=0)$. One has $(n-C_0)/C_0 = (n(\Delta)-n(\Delta=0))/n(\Delta=0)\sim 10^{-5}$, so that the reduction of the angular momentum is crucial. 

To connect the hydrodynamic anomalies in Eqs.(\ref{Current})-(\ref{L}) and the chiral anomaly in relativistic theories, let us take into account that the parameter $k_F$ which enters $C_0$ marks the position of the Weyl points in $^3$He-A:
${\bf K}_a=\pm k_F\hat{\bf l}$. When ${\bf K}_a\rightarrow 0$, the Weyl points merge and annihilate, as a results the hydrodynamic anomalies disappear. They do not exist in the strong coupling regime, where the chiral superfluid has no Weyl points and $C_0=0$.  All the dynamic anomalies experienced by $^3$He-A result from the existence of the Weyl points, which allow the spectral flow from the vacuum. The state of the chiral superfluid with anomalies and the anomaly-free state of the chiral superfluid are separated by the topological quantum phase transition.

Close to the Weyl point the spectral flow can be considered in terms of the relativistic fermions. The expansion of the Hamiltonian (\ref{H}) in the vicinity of the Weyl point produces the following  relativistic Hamiltonian describing the chiral Weyl fermions: 
\begin{equation}
H({\bf k})=e_\alpha^i\tau^\alpha \left(p_i-qA_i\right) \,.
\label{SU2}
\end{equation}
Here 
${\mbox{\boldmath$\tau$}}$ are Pauli matrices corresponding to the Bogoliubov-Nambu  isotopic spin; $q{\bf A}$ is the position of the Weyl point ${\bf K}_\pm$ which in the inhomogeneous $^3$He-A corresponds to the effective gauge field ${\bf A}({\bf r})= k_F\hat{\bf l}({\bf r})$ acting on the effective electric charge $q=\pm 1$ of the Weyl fermions.
The matrix $e_\alpha^i$ describes the effective (synthetic) tetrad field with ${\bf e}_1= (\Delta/k_F)\hat{\bf m}$, ${\bf e}_2= (\Delta/k_F)\hat{\bf n}$ and
${\bf e}_3=\pm (k_F/m)\hat{\bf l}$. The determinant of the matrix determines the chirality of the Weyl fermions. Here we ignore the spin degrees of freedom, which for the inhomogeneous $\hat{\bf d}$-field gives rise to the synthetic $SU(2)$ field in addition to the Abelian  field ${\bf A}({\bf r},t)$.\cite{Volovik2003}

The chiral fermions experience the effect of chiral anomaly in the presence of the synthetic 
electric and magnetic fields 
\begin{equation}
{\bf E}=-k_F\partial_t\hat{\bf l} ~~,~~{\bf B}=k_F\nabla \times \hat{\bf l} 
\,.
\label{EMfields}
\end{equation}
The fermions created from the superfluid vacuum carry the fermionic charge from the vacuum to the "matter" -- the normal component of the liquid, which at low temperatures consists of thermal Weyl fermions. For us the important fermionic charge is the quasiparticle momentum: each fermion created from the vacuum carry with it the momentum ${\bf K}^{(a)}= \pm k_F\hat{\bf l}$. According to the Adler-Bell-Jackiw equations for chiral anomaly, this gives the following momentum creation from the vacuum per unit time per unit volume:
\begin{equation}
\partial_t P_i -\nabla_k\pi_{ik} =\frac{1}{4\pi^2}  {\bf B} ({\bf r},t)\cdot {\bf E} ({\bf r},t)\sum_a   K_i^{(a)} N_a q_a^2 
\,.
\label{MomentumProductionGeneral}
\end{equation}
Since in the supefluids the momentum density equals the mass current density, ${\bf P}={\bf j}$, the Eq.(\ref{MomentumProductionGeneral}) reproduces the equation (\ref{nonconservation}). This demonstrates that nonconservation of the linear momentum of the vacuum is the consequence of the chiral anomaly.

\begin{figure}
 \includegraphics[width=0.5\textwidth]{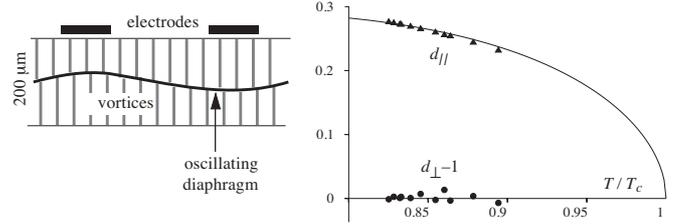}
 \caption{
Fig. 3. Experimental verification of the Adler-Bell-Jackiw anomaly equation in $^3$He-A
 (from Ref.\cite{BevanNature1997}).
{\it Left}: A uniform array of vortices is produced
by rotating the whole cryostat, and oscillatory superflow perpendicular
to the rotation axis is produced by a vibrating diaphragm, while
the normal fluid (thermal excitations) is clamped by viscosity, ${\bf v}_{\rm
n}=0$. The velocity
${\bf v}_{\rm L}$ of the vortex array is determined by the overall
balance of forces acting on the vortices in Eq.(\ref{ForceBalance}).  The
vortices
produce the dissipation $d_\parallel$ and also the
coupling between two orthogonal modes, which is proportional to
$d_\perp$.  {\it Right}: The parameters $d_\parallel$ and $d_\perp$ measured for  the continuous vortices-skyrmions  in $^3$He-A. As distinct from the $^3$He-B vortices, for skyrmions the measured parameter  $d_\perp$ is close to unity. According to Eq.(\ref{d_perpAphase}), the experiment demonstrates that the anomaly parameter $C_0$ is close to the particle density $n$ and thus it verifies the Adler-Bell-Jackiw anomaly equation (\ref{MomentumProductionGeneral}).
 }
 \label{Bevan}
\end{figure}

The creation of the momentum from the vacuum per unit time due to the spectral flow through the Weyl nodes under the effective electric and magnetic fields produced by  the time dependent texture of the vector $\hat{\bf l}({\bf r},t)$ according to Eq.(\ref{EMfields}), means that there is an extra force acting on the texture. The relevant  time dependent texture 
is the moving vortex-skyrmion, where 
$\hat{\bf l}({\bf r},t)=\hat{\bf l}({\bf r} -{\bf v}_{\rm L} t)$. The anomalous spectral-flow force acting on this topological object moving with velocity ${\bf v}_{\rm L}$ is obtained after integration of the rhs of Eq.(\ref{MomentumProductionGeneral}) over the cross-section of the skyrmion:
\begin{equation}
{\bf F}_{\rm ~spectral~flow}= - \pi {\cal N} C_0 {\hat {\bf z}}  \times {\bf v}_{\rm L} \,.
\label{KopninForce}
\end{equation} 
Here ${\cal N}$ is the vortex winding number of the texture, which via Eq.(\ref{Combined}) 
is expressed in terms of the Weyl point charges $N_a$ and  the topological $\pi_2$ charges of their spatial distributions in the texture. This demonstrates that the spectral flow force acting on
the vortex-skyrmion, which has ${\cal N}=2$, is the result of the combined effect of real-space and momentum-space topologies.

The force acting on the vortex-skyrmions has been measured experimentally.\cite{BevanNature1997}
There are several forces acting on vortices, including the Magnus force, Iodanski force  and the anomalous spectral-flow force, which is called the Kopnin force. For the steady state motion of vortices the sum of all forces
acting on the vortex must be zero. This gives the following equation connecting velocity of the
superfluid vacuum ${\bf v}_{\rm s}$, the velocity of the vortex line ${\bf v}_{\rm L}$ and the velocity ${\bf v}_{\rm n}$ of "matter" -- the velocity of the normal component of the liquid:
\begin{equation}
\hat{\bf z}\times ({\bf v}_{\rm L}-{\bf v}_{\rm s})+
d_\perp \hat{\bf z}\times({\bf
v}_{\rm n}-{\bf v}_{\rm L})+d_\parallel({\bf v}_{\rm n}-{\bf v}_{\rm L})
=0\,.
\label{ForceBalance}
\end{equation}
For the continuous vortex-skyrmion in $^3$He-A with the spectral flow force in Eq.(\ref{KopninForce}) the reactive parameter $d_\perp$ is expressed in terms of the anomaly parameter $C_0$:
 \begin{equation}
d_\perp-1= \frac{C_0-n}{n_{\rm s}(T)}\,.
\label{d_perpAphase}
\end{equation}
Here $n_{\rm s}(T)=n- n_{\rm n}(T)$ is the density of the superfluid component.

Since in $^3$He-A the anomaly parameter $C_0$ is very close to the particle density $n$, the chiral anomaly in $^3$He-A should lead to equation $d_\perp-1=0$ for practically all temperatures. This is what has been observed in Manchester experiment on skyrmions in $^3$He-A, 
see Fig. \ref{Bevan} ({\it right}) which experimentally confirms
the generalized Adler-Bell-Jackiw equation (\ref{MomentumProductionGeneral}). 

In conclusion, the chiral anomaly related to the Weyl fermionic quasiparticles, whose gapless spectrum is protected by the topological invariant in ${\bf k}$-space, has been observed in the experiments with skyrmions --  objects, which are protected by the topological invariant in the ${\bf r}$-space.
The effect of chiral anomaly observed in $^3$He-A incorporates several topological charges described by the combined topology in the extended $({\bf k},{\bf r})$-space, which is beyond the conventional anomalies in the relativistic systems. 

There are the other consequences of chiral anomaly in $^3$He-A, such as the chiral magnetic effect.
In particle physics  the chiral magnetic effect is the appearance  of the non-dissipative current along the magnetic field due to the chirality imbalance. It is studied in relativistic heavy ion collisions where strong
magnetic fields are created by the colliding ions, see review \cite{Kharzeev2015}.
This effect is encoded in the Chern-Simons term in the free energy
\cite{Volovik2003}
\begin{equation}
F_{\rm CS}=- \frac{1}{8\pi^2} \int d^3r  {\bf A}\cdot ({\bf\nabla}\times {\bf A}) \sum_a N_a \mu_a
q_a^2 \,.
\label{ChernSimons}
\end{equation}
The variation $\delta  F_{\rm CS}/\delta {\bf A}$ gives the current ${\bf J}$ along magnetic field ${\bf B}$. 

One may argue that the chiral magnetic effect in the ground state is prohibited by 
the Bloch theorem \cite{Yamamoto2015}.
But in our case the field ${\bf A} =  k_F\hat{\bf l}$ is effective, and the corresponding current 
${\bf J}=\delta  F/\delta {\bf A}$ is also effective.  It does not coincide with the real mass current 
${\bf j}=\delta  F/\delta {\bf v}_{\rm s}$. The imbalance between the chiral chemical potentials of left-handed and right-handed Weyl fermions is also effective: it is provided by the counterflow due to the Doppler shift: 
$\mu_\pm =\pm k_F \hat{\bf l}\cdot ({\bf v}_{\rm n}-{\bf v}_{\rm s})$.  For the effective fields and currents the no-go theorem is not applicable, and the Chern-Simons term (\ref{ChernSimons}) gives rise to the helical instabilty of the counterflow, which  has been experimentally 
confirmed \cite{ExperimentalMagnetogenesis1997}. 

The spectral flow may also occur through the nodes in the spectrum of the edge states.  The spectral flow through the nodes in bulk \cite{Volovik1995} and on the surface \cite{Tada2015,Volovik2015} leads to the spectral asymmetry, which is responsible for the anomaly related to the magnitude of the internal angular momentum in chiral superfluids.

We considered the effects of combined topology,  which connects the Weyl points in ${\bf k}$-space
and vortices and skyrmions in ${\bf r}$-space. The consideration can be extended to the more complicated topological objects, such as nexus in real space \cite{Cornwall1999}  and nexus in momentum space \cite{TeroNexus2015}, where one may expect exotic consequences of the spectral flow.

This work has been supported in part by the Academy of Finland (project no. 284594),
and by the facilities of the Cryohall infrastructure of Aalto University.

\end{document}